\documentclass[preprint,12pt]{elsarticle}

\usepackage{graphicx}

\usepackage{amssymb}

\journal{Nuclear Instruments and Methods in Physics Research A}

\begin{document}

\begin{frontmatter}

  \title{Evaluation of the discovery potential of an underwater Mediterranean neutrino telescope taking into account the estimated directional resolution and energy of the reconstructed tracks}

 \author{A.~Leisos\corref{cor1}}
  \cortext[cor1]{Corresponding author.~Tel: +302610367523; Fax: +302610367528}
  \ead{leisos@eap.gr}
  \author{A.G.~Tsirigotis}
  \author{and S.E.~Tzamarias}
  \author{on behalf of the KM3NeT consortium}
  \address{Physics Laboratory, School of Science and Technology, Hellenic Open University, Tsamadou 13-15 and Ag.~Andreou, Patras 26222, Greece}

\begin{abstract}
We report on the development of search methods for point-like and extended neutrino sources, utilizing the tracking and energy estimation capabilities of an underwater, Very Large Volume Neutrino Telescope (VLVnT). We demonstrate that the developed techniques offer a significant improvement on the telescope's discovery potential. We also present results on the potential of the Mediterranean KM3NeT to discover galactic neutrino sources.
\end{abstract}


\end{frontmatter}
\section{Introduction}
The geographical position of the Mediterranean is ideal, since the area of the sky that is possible to be observed by a deep-sea neutrino telescope will include the centre of our galaxy and most of the galactic plane where many high-energy gamma-ray sources (potential neutrino emitters) have been discovered. The optical properties of water and the geographical position could elevate a Mediterranean VLVnT, KM3NeT, to be the most sensitive cosmic neutrino telescope. 
In optimizing the KM3NeT configuration, it is important to utilize the maximum of the experimental information in order to maximize the discovery potential of the telescope.  In this work, we demonstrate that by employing in the search strategies information based on the directional resolution and the measured energy of each of the detected muons, the discovery potential of the telescope is significantly improved. 
In Section 2 we describe the detector configuration and we highlight the performance of the HOURS \cite{hours-1} software package to simulate the response of the detector and to reconstruct the muon-neutrino and muon-antineutrino\footnote{Hereafter, we use the term ``neutrino'' for both muon-neutrinos and muon-antineutrinos.} direction from the experimental signal. In Section 3 we describe the analysis strategy for point-like and extended (disc-type or Gaussian-type) sources. In Section 4 we present results in estimating the KM3NeT discovery potential for galactic sources whilst Section 5 summarizes the results of our study.
\section{Detector Configuration and Signal Reconstruction}
In this study we assumed that the KM3NeT telescope will consist of 308 vertical structures (Detection Units - DUs), deployed at a Mediterranean site ($37^\circ$ north and $18^\circ$ east) in 3500 m depth. A DU carries photo-sensors arranged vertically on 20 horizontal bars (floors), with a distance between consecutive floors of 40~m, as described in \cite{TDR}. The DUs  fill a cylindrical volume of 5.8 $\mathrm{ km^3}$, 1560~m in radius and 760~m in height\footnote{For some special studies in this work we considered the half of the detector, consisting of 154 DUs in a cylindrical volume of 2.9 $\mathrm{km^3}$, with the same DU density as the whole telescope.}. The positions of the DUs  on the seabed are almost-uniformly distributed with a typical distance between DUs 180~m. Each horizontal bar of a DU, at its end, supports two photo-sensor units, hereafter called Digital Optical Modules (DOMs). A DOM consists of a 17-inch diameter pressure resistant glass sphere housing 31 3-inch photomultiplier (PMT) tubes, their high-voltage bases and their interfaces to the data acquisition system with nanosecond timing precision. The front-end electronics perform an accurate measurement of the PMT-signal arrival time and an estimation of the signal amplitude through a Time Over Threshold (TOT) measurement \cite{km3netppm}.

We have simulated the response of this telescope configuration to 15 million neutrinos with directions uniformly distributed on the sky, with energies between 100~GeV to 100~PeV, following an energy spectrum proportional to $E^{-2}$. In parallel we simulated, using CORSIKA \cite{corsica}, the telescope's response to muons and muon-bundles produced in 15 million down-going Extensive Air Showers (EAS). The simulated EAS have energies between 1~TeV and  1~EeV, following the experimentally established \cite{grieder}  energy distribution and composition of the charged cosmic rays. The direction of the primary cosmic ray on the upper atmosphere is distributed uniformly on the sky up to an angle of 3 degrees above the detector horizon. We have used HOURS to simulate all the relevant physical processes, the production of signal and background, the response of the PMTs as well as the functionality of the digitization electronics. HOURS also provides several data analysis tools that have been used to reconstruct the direction (estimating also accurately the relevant error matrix of the track parameters) of the muons  as well as  to estimate the muon's energy.
We have tuned the tracking quality criteria in order to minimize the probability to miss-reconstruct down-going atmospheric muons (or muon bundles) as up-going products of neutrino interactions.  The method, used in HOURS, to estimate the energy of a reconstructed muon, is described in \cite{hours-2}. For this telescope configuration, the log-energy resolution, $\Delta \left ( \log E \right )$, is described by the empirical formula $$\Delta \left ( \log E \right )=0.32+0.082\times \arctan \left ( 10-2.42\times \log E \right )$$
where $E$ is the true muon energy in GeV at the closest distance of approach to the center of the detector. The log-energy resolution varies from 0.42 at 1 TeV to 0.23 at 100 TeV, reaching the  value 0.2  at 1 PeV. 
\section{Analysis Strategy}
As in \cite{neun} we choose, for each reconstructed track, a suitable coordinate system (hereafter Track Reference Frame - TRF) in which
the direction of the parent-neutrino with respect to the reconstructed muon direction is described by the uncorrelated orthogonal coordinates $\theta$ and $\phi$. 
In this system the angle $\psi$ between the parent-neutrino and the reconstructed muon is very well approximated (for small values of $\psi$) as $ \psi^2 =\theta ^{2}+\phi^{2}$. 

Restricting the analysis to events where the reconstructed muon deviates less than $R_{\mathrm{max}}$  from the parent-neutrino direction (i.e.~$\theta^{2}+\phi^{2} \leq R^{2}_{\mathrm{max}}$) the probability density function (PDF) of $\theta$ and $\phi$ 
has the following Gaussian form:
\begin{equation}
P_{\mathrm{ang}}\left (\theta ,\phi   \right ) =\Re \times \frac{1}{2\pi \times \Sigma _{\mathrm{\theta} }\times \Sigma_{\mathrm{\phi}}}\times e^{-\frac{1}{2}\left ( {\frac{\mathrm{\theta ^{2}}}{\Sigma _{\mathrm{\theta }}^{2}} + \frac{\mathrm{\phi} ^{2}}{\Sigma _{\mathrm{\phi  }}^{2}}    } \right)} 
\label{1}
\end{equation}
where the normalization factor $\Re$ is  well approximated as
$$\Re=\left (1-e^{-\frac{R^{2}_{\mathrm{max}}}{\Sigma_{\theta}^{2}+\Sigma_{\phi}^{2}  } } \right )^{-1}$$

The angle between the parent-neutrino and the reconstructed muon, is due to the kinematics of the parent-neutrino interaction as well as
to the limited detector resolution in estimating the muon track. Consequently,  
the values of the sigma parameters, $\Sigma_{\theta}$ and $\Sigma_{\phi}$ in (\ref{1}), are evaluated as convolutions of the error matrix elements $\sigma_{\theta}$ and $\sigma_{\phi}$ of the reconstructed muon track due to the detector resolution, with the  RMS of the misalignments in $\theta$ and $\phi$ of the parent-neutrino with the true product-muon momentum, $s_\theta$ and $s_\phi$ respectively (i.e.~$\Sigma _{\theta }^{2}=\sigma _{\theta }^{2}+s^2_\theta$ and $ \Sigma _{\phi }^{2}=\sigma _{\phi }^{2}+s^2_\phi$ ). 
However, due to the symmetry of the parent-neutrino direction around the muon track, it follows that $s_{\theta}=s_{\phi}=s$ and since the only available experimental information relative to the parent-neutrino energy is the estimated energy \cite{hours-2} of the reconstructed muon, $E_m$, we used  the simulated set of events described in Section 2, to parametrize this misalignment as a function of $E_m$  ($s_{\theta}\left ( E_m \right )= s_{\phi}\left ( E_m \right )= s\left ( E_m \right )$).
 
When a reconstructed muon track points inside a ring of angular radius $R_{\mathrm{max}}$ around  a point source, Eq.~(\ref{1}) expresses  the signal-PDF of the track to be produced from a neutrino emitted by the source. Restricting the analysis only to up-going reconstructed muons and having eliminated the background due to miss-reconstructed atmospheric muons, the only background source is the atmospheric neutrinos. For small values of $R_{\mathrm{max}}$\footnote{$R_{\mathrm{max}}$ is chosen large enough to include almost all the muon tracks from the point source e.g.~of the order of few degrees.} the PDF describing the directional distribution of the background muons around the source direction is constant over the solid angle:
\begin{equation}
\Pi\left ( R_{\mathrm{max}} \right )=\left(\Delta \Omega \right)^{-1}\simeq \left(\pi R_{\mathrm{max}}^2 \right)^{-1}
\label{3}
\end{equation}

\begin{figure}
\centering
\includegraphics*[width=12.0cm]{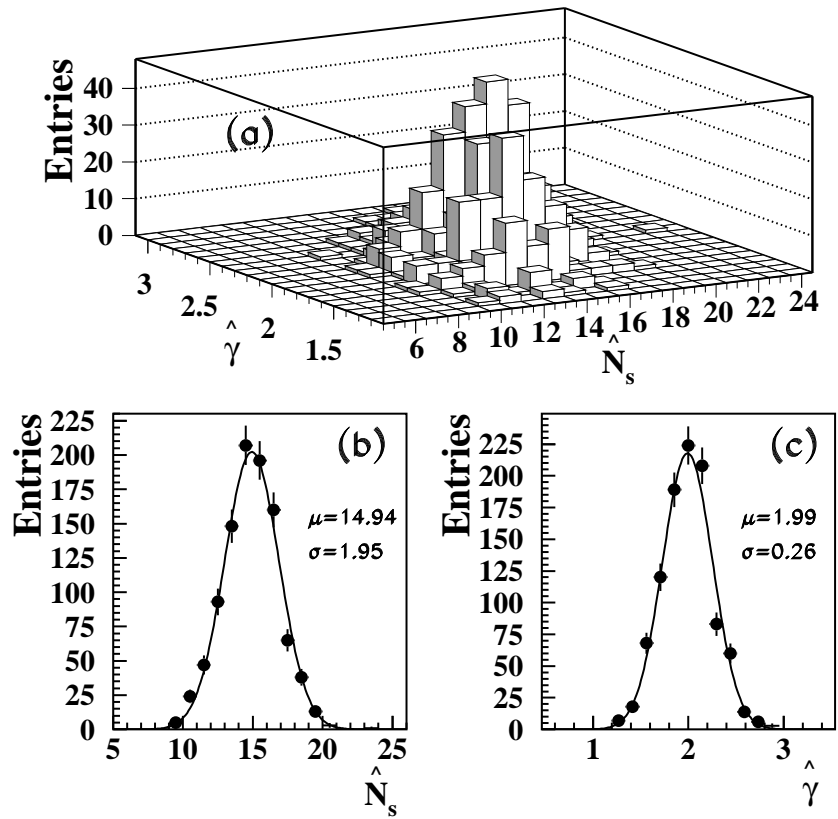}
\caption{Results of 1000 simulated experiments (see text). In each experiment the number of the signal events ($N_s$) and the spectral index ($\gamma$) of the source were simultaneously estimated. (a) Two-dimensional distribution of the estimations $\hat{N_s}$ and $\hat{\gamma}$. (b and c) Distributions of the estimated values of $\hat{N_s}$ and $\hat{\gamma}$ respectively (data points) fitted by gaussian functions (solid lines). The mean ($\mu$) and sigma ($\sigma$) values of the gaussians are also shown.}
\label{fig_fit}
\end{figure}

In general, the energy spectra of signal and background neutrinos are different, resulting thus in different measured-energy distributions of the corresponding muons. We have used samples of simulated events, properly reweighted according to the energy-zenith differential distribution of atmospheric neutrinos \cite{bartol}, in order to evaluate the distribution of measured energies of the background in bins of the cosine of the zenith\footnote{The zenith angle refers to to the detector reference frame where Z is parallel to the zenith axis and the XY plane is parallel to the horizon.} angle, $\theta_{z}$. Similarly, by reweighting the simulated events, we were able to evaluate the distribution of measured energies of muons produced by neutrinos from a source, which is located at any declination $\delta$ and emits neutrinos with a certain energy spectrum. In order to take advantage of both, the good pointing resolution and the energy information provided by the underwater telescope, we define the signal PDF, $P^{s}\left(\theta ,\phi, E_m ; \gamma \right)$, and the background PDF, $P^{b} \left( E_{m} \right)$,  as:
\begin{equation}
P^{s}\left(\theta ,\phi, E_m ; \gamma \right) = P^{s}_{\mathrm{ang}}\left(\theta ,\phi \right) \times  P^{s}_{\mathrm{en}}\left ( E_{m};\gamma,\mathrm{cos}\theta_{z} \right ) 
\label{4}
\end{equation}
\begin{equation}
P^{b} \left( E_{m}\right)=\Pi \left( R_{\mathrm{max}} \right) \times  P^{b}_{\mathrm{en}}\left ( E_{m};\mathrm{cos}\theta_{z} \right ) 
\label{4b}
\end{equation}
where $P^{s}_{\mathrm{en}}\left ( E_{m};\gamma,\mathrm{cos}\theta_{z} \right )$  is the signal PDF of the measured-energy, evaluated in 20 bins of $\mathrm{cos}\theta_{z}$  (between -1 and 1) by reweighting the simulated events to correspond to a $E^{-\gamma}$ neutrino spectrum, and $P^{b}_{\mathrm{en}}\left ( E_{m};\mathrm{cos}\theta_{z} \right )$ is the background PDF of the measured-energy, evaluated in the same bins of $\mathrm{cos}\theta_{z}$. Furthermore, assuming that $\alpha_{s}$ is the fraction of the signal events in the sample of $N$ selected muon tracks, we can write the Likelihood function as:
\begin{eqnarray}
 L \left ( \alpha_{s},\gamma \right)=  \prod_{i=1,N} \{ \alpha_{s} \times P^{s}\left(\theta^i ,\phi^i, E^{i}_m ; \gamma \right) \\
+ \left( 1- \alpha_{s} \right) \times P^{b} \left( E^{i}_{m}\right)\} \nonumber
\label{5}
\end{eqnarray}
which can be used to estimate both the signal fraction and the spectral index of the source. 
As a demonstration, we performed 1000 experiments using simulated set of events and estimating simultaneously the $\alpha_{s}$ and $\gamma$ parameters by maximizing the logarithm of Eq.~(5) in each set. Each data set consisted: a) of 15, on average, background reconstructed muons  with directions uniformly distributed inside an angular ring of radius $R_{\mathrm{max}}=0.6^\circ$ around a hypothetical point-like neutrino source at $\delta =-60^\circ$ and b) of 15 more reconstructed muons  produced by neutrinos from the source with an energy spectrum proportional to $E^{-2}$ (i.e.~$\gamma=2$). The estimations are presented in Fig.~\ref{fig_fit}, demonstrating a very good agreement with the true parameter  values. 
As discussed in \cite{braun}, the likelihood estimator, expressed as in (5), can result in negative values of $\hat{\alpha_s}$\footnote{Hereafter, we use the symbol `` $\hat{}$ `` over a variable to indicate the result of an estimation (e.g.~$\hat{x}$ is an estimation of the variable $x$).} in cases that the background is much higher than the signal.   We have used  the  expected number of background events, $B$, in the sample\footnote{This  expected background  can be evaluated accurately with simulation studies and even more accurately by reshuffling the azimuth of the reconstructed muons in real data as in \cite{icecube}.} to define the following extended likelihood function:
\begin{equation}
L_{\mathrm{ext}} \left ( \alpha_{s}, \gamma \right)= \frac{{\mu}^{N}\times e^{-\mu}}{N!}  \times  L \left ( \alpha_{s} ,\gamma \right)
\label{6}
\end{equation}
where   $\mu=B/ \left(1- \alpha_{s}\right)$ is the expected number of total events in the sample, $\alpha_s$ is the signal fraction and  $B$ the mean number of background events. The extended likelihood function (\ref{6}) gives systematically a slightly better estimation resolution than (5) and the fit procedure converges faster. Following the above definitions, $\mu$ must be greater or equal to $B$ and consequently $a_s$ varies in the interval $[0,1]$.

Using (\ref{6}), a resolution parameter $\lambda$ is defined as the following log-likelihood ratio:
\begin{equation}
\lambda=-2 \times \ln \frac{ L_{\mathrm{ext}} \left( \alpha_{s}=0\right )}{L_{\mathrm{ext}} \left( \hat{\alpha_{s}},\gamma=\gamma_{exp}\right )}
\label{7}
\end{equation}
where the numerator corresponds to the hypothesis that all the tracks in the selected sample of events are products of atmospheric-neutrino interactions and the denominator is the maximum likelihood value  under the hypothesis that the energy distribution of the reconstructed muons corresponds to a known (expected) energy spectrum of the parent-neutrinos (i.e.~in the fit we estimate only $\alpha_{s}$). 
The discovery potential is defined as the minimum signal flux that has $50\%$ probability to be detected during a certain observation time (usually one year) with at least 5 sigma significance (p-value=$2.86 \times 10^{-7}$ in the $\lambda$ distribution) above the background.
For any source, the discovery potential of a neutrino telescope is evaluated by utilizing Monte Carlo experiments and comparing the distribution of the $\lambda$  parameter corresponding to background-only events with distributions of $\lambda$  corresponding to sets comprising signal and background as described in \cite{braun,icecube}.
 
In the rest of this Section we demonstrate the performance of this technique (hereafter fit-technique) to search for point and extended neutrino sources with half of the neutrino telescope (i.e.~154 DUs) described in Section 2. In the case of a hypothetical point source, located at $\delta=-60^\circ$ that emits neutrinos with an energy spectrum proportional to $E^{-2}$ and using a search ring of $R_{\mathrm{max}}=0.6^\circ$, the 5$\sigma$ discovery potential was found to be $2.80\times 10^{-9}$ $ \left(E/\mathrm{GeV}\right)^{-2}$ $ \mathrm{GeV^{-1}s^{-1}cm^{-2}}$. We have also applied to the same samples of simulated events the so called bin-technique \cite{TDR,apostol}, which counts tracks pointing inside an angular ring and compares their number with the expected number of background muons, without taking into account the pointing resolution and the measured energy. The bin-technique, after optimizing the radius of the search ring, leads to an inferior discovery potential: $3.80\times 10^{-9}$ $ \left(E/\mathrm{GeV}\right)^{-2}$ $\mathrm{GeV^{-1}s^{-1}cm^{-2}}$. \\
The gain in sensitivity (observed also in \cite{icecube}) is a result of the extra information, on the tracking and energy resolution of the detector, which is included on a track by track basis. It must be also noticed that the results of the bin-method strongly depend on the size of the search ring whilst the fit-technique is almost insensitive to such a choice, as long as $R_{\mathrm{max}}$ is greater than 2 or 3 times the median of the angle between the parent-neutrino and the reconstructed muon direction. However, there are variations of the bin-technique where the discovery potential is evaluated by tuning simultaneously the reconstruction (e.g.~criteria concerning the selection of the ``best track solution'') and the track selection criteria (blindly) in order to achieve optimum results. Although,  this blind optimization leads in some cases to similar results as the fit-technique (e.g.~the blindly optimized bin-technique finds for this source and detector configuration $2.85\times 10^{-9} $ $\left(E/\mathrm{GeV}\right)^{-2} $ $\mathrm{GeV^{-1}s^{-1}}$ $\mathrm{cm^{-2}}$) such a method is not bias-free when applied to experimental data. 

In order to quantify the contribution of the energy information to the sensitivity of the fit-technique, we evaluated the discovery potential by employing only the tracking information to form the extended likelihood function, i.e.~$P^{s} \left(\theta,\phi;\gamma \right)= P^s_{\mathrm{ang}}\left(\theta,\phi \right)$  and $P^{b}=\Pi\left(R_{\mathrm{max}}\right)$. In this case the discovery potential is found to be $3.20\times 10^{-9} $ $\left(E/\mathrm{GeV}\right)^{-2}$ $\mathrm{GeV^{-1}s^{-1}cm^{-2}}$, which when compared  with the above results indicates that 40$\%$ of the gain in sensitivity offered by the fit-technique (for $E^{-2}$ neutrino spectrum) is due to the energy information. 

Most of the galactic neutrino sources candidates are extended objects. In case that the source has a Gaussian shape as  several of the H.E.S.S. sources (e.g.~HESSJ1616-508 or HESSJ1614-518) \cite{hess-source}, the fit-technique is easily modified to account for the source shape by incorporating the source's Gaussian sigma, $\rho$, to  Eq.~(\ref{1}) as:
$\Sigma _{\theta }^{2}=\sigma _{\theta }^{2}+s^2\left ( E_m \right )+\rho^2$ and $ \Sigma _{\phi }^{2}=\sigma _{\phi }^{2}+s^2\left ( E_m \right )+\rho^2$. However, in the case that the source is a uniformly emitting disc of radius $d$, Eq.~(\ref{1}) should be integrated over the whole area of this disc.

Due to the fact that such an integration is very computer-time consuming, an independent set of simulated neutrino-induced muon tracks was used to parametrize the expected angular profile of the reconstructed muons around the center of the disc. The angular signal-PDF, $P^s_{\mathrm{ang}}\left (\theta ,\phi   \right )$ is replaced in the extended likelihood function (Eqs. \ref{3}, \ref{4} and \ref{6}), by $P_{\mathrm{profile}}^{\mathrm{disc}}\left( r/d;\mathrm{cos}\theta_{z} \right)$, which expresses the PDF of  a  reconstructed (signal) muon to deviate by an angle $r$ from the direction of the center of the source-disc (of radius $d$), when the zenith angle of the reconstructed muon is $\theta_{z}$.  We used the fit-technique with this modification to evaluate the discovery potential of the half detector for an extended source at $\delta=-60^\circ$ with radius $d=0.6^\circ$, emitting neutrinos with an energy spectrum proportional to  $E^{-2}$. Using a search ring with $R_{\mathrm{max}}=1.2^\circ$ around the center of the disc, we found the discovery potential to be:
 $4.65\times 10^{-9} $ $\left(E/\mathrm{GeV}\right)^{-2}$ $\mathrm{GeV^{-1}s^{-1}cm^{-2}}$.\\
 This result is by 67\% worse than the discovery potential corresponding to a point source at the same declination, reflecting the fact that the source extension  weakens the impact of the angular resolution to the sensitivity of the detector for such sources. However, even in the case of this extended source, the performance of the bin-technique  ($5.60\times 10^{-9} \left(E/\mathrm{GeV}\right)^{-2}\mathrm{GeV^{-1}s^{-1}cm^{-2}}$)  is inferior compared to the results of the fit-technique.  
\section{Discovery Potential for Galactic Sources}
In this Section we study the potential of the whole KM3NeT detector (308 DUs) to discover neutrino emission from galactic objects that have been identified as sources of TeV gamma rays. We focus on the bright gamma ray source RXJ1713.7-3946 \cite{rxj17}, assuming that the observed Very High Energy (VHE) gamma rays emitted from this object are of hadronic origin \cite{hadr} and that the expected neutrino flux (hereafter reference flux) is described as: $1.68\times10^{-11} E^{-1.72} e^{-\sqrt{E/2.1}}~\left(\mathrm{TeV^{-1} cm^{-2} s^{-1}}\right)$ (where E is in TeV). We modeled this source as a radiating disc of angular radius of $d=0.6^\circ$ centered at $\delta=-39^\circ 46'$ and we evaluated the discovery potential by the fit-technique, reweighting the simulated events according to the reference energy spectrum. The 5$\sigma$ discovery potential for one year observation  was found to be 3.75 times the reference flux when the energy information is included in the likelihood function (3.9 times the reference flux without including the energy information in the fit).  The bin-technique, optimized for best results, offers an inferior discovery potential,  4.6 times the reference flux \cite{apostol}. 
However, it is important to estimate the observation time required to achieve a 5$\sigma$ discovery (hereafter 5$\sigma$-Discovery Time, $5\sigma-DT$) in the case that RXJ1713.7-3946 emits neutrinos according to the expected flux. Using the fit-technique we evaluated the 5$\sigma$ (3$\sigma$) discovery potentials for several observation times ranging from 1 to 15 years. The  $5\sigma-DT$ ($3\sigma-DT$) is the observation time that corresponds to a discovery potential equal to the reference flux. When the energy information is included the $5\sigma-DT$  is 7.5 yr, whilst the $5\sigma-DT$ increases to 8 yr in the case that only the angular information is used in the fit (in comparison the bin-technique, blindly optimized for best results, offers  a $5\sigma-DT$ of 13.1 yr). We have also found that the observation time required for a $3\sigma$ discovery ($3\sigma-DT$) with this detector configuration is 3 yr (with energy information included) and 4 yr (without the energy information). The statistical errors on the above estimates of the discovery time are of the order of 1 yr for the  $5\sigma-DT$ and 0.5 yr for the $3\sigma-DT$.
\begin{table*}[t]
  \centering
\caption{Estimation of the discovery potential of KM3NeT  and of the required observation times for $5\sigma$ and $3\sigma$ discoveries of candidate galactic sources. The discovery potentials are estimated for one year of observation and they are expressed  as multiples of the reference flux of each source. The discovery times are in years and the errors reflect the estimated statistical uncertainties.}
\scriptsize
{
\hfill{}
\begin{center}
  \begin{tabular}{ c c c c c c }
    \hline
Candidate        & Reference Flux                               & $5\sigma$ (1 yr)  & $5\sigma$       & $3\sigma$ (1 yr)    &  $3\sigma$      \\ 
Source           & ($\mathrm{TeV^{-1} cm^{-2} s^{-1}}$)            & Discovery         & Discovery       & Discovery           &  Discovery      \\
                 &                                              & Potential         & Time            & Potential           & Time            \\ \hline
RXJ1713.7-3946   & $1.68\times 10^{-11}  E^{-1.72} e^{-\sqrt{E/2.1}}$   & 3.75              & 7.5$\pm1.0$     &   2.14              &  3.0$\pm0.5$   \\ 
RXJ0852.0-4622   & $1.676\times 10^{-11} E^{-1.78} e^{-\sqrt{E/1.19}}$  & 6.30              & 33$\pm4$        &   3.71              & 11.0$\pm0.5$    \\ 
HESSJ1614-518    & $0.46\times 10^{-11}  E^{-2.2} e^{-\sqrt{E/2.1}}$    & 16.15            & -               &   8.85              &  -             \\ 
HESSJ1616-508    & $0.42\times 10^{-11}  E^{-1.72} e^{-\sqrt{E/2.1}}$   & 9.54              & -               &   5.11              &  28$\pm4$       \\ 
\end{tabular}
\end{center}
}         
\hfill{}
\label{d5rxj}
\end{table*}

We have applied the fit-technique to evaluate the potential of the detector to discover other fainter (potential) neutrino sources, which have been observed to emit VHE gamma rays, assuming that the observed TeV gamma spectrum is of hadronic origin. For example we studied the case of  RXJ0852.0-4622\cite{rxj08} (disc with a radius of $1^\circ$), HESSJ1614-518 \cite{hess1614} (Gaussian shape with $\sigma=0.25^\circ$) and HESSJ1616-508 \cite{hess1614} (Gaussian shape with $\sigma=0.16^\circ$). The prediction of \cite{kappes} has been used as a reference flux for  RXJ0852.0-4622. For the other two sources we imposed an energy cutoff (the same as  RXJ1713.7-3946) to the neutrino spectra by fitting the corresponding predictions of \cite{kappes} (which is of the form $E^{-\gamma}$ due to the limited statistics of the observed multi-TeV gamma rays) to the general form $N\times  E^{-s} e^{-\sqrt{E/2.1}}$ with free parameters $N$ and $s$ for each source. 
We have found the discovery potentials of this detector configuration for one year of observation time as well as the discovery times in years. The results are summarized in Table 1.
\section{Conclusions}
We have applied the fit-technique in order to estimate the potential of a large underwater Mediterranean neutrino telescope to discover point-like, (extended) Gaussian-like and (extended) disc-like neutrino sources. We have demonstrated that the pointing and energy information offered by this detector, when taken into account on a track-by-track basis, enhances significantly the discovery potential of the telescope. We have estimated the 5$\sigma$ discovery potential of the telescope for the neutrino emission from the Supernova remnant RXJ1713.7-3946 under the assumption that the observed VHE gamma spectrum of this source is of hadronic origin. The required flux for 5$\sigma$ discovery in 1 yr of data taking is:\\
 $3.75 \times \{1.68\times10^{-11} E^{-1.72} e^{-\sqrt{E/2.1}}~\left(\mathrm{TeV^{-1} cm^{-2} s^{-1}}\right)\}$.\\

We have also estimated that 7.5 yr of observation are required to establish a $5\sigma$  discovery of this source (in comparison with  13.1y in case the bin-technique is used).
We have also studied the potential of this telescope to discover other candidate neutrino sources in the galactic plane and we found that several decades of observation are required. However, it must be emphasized that in this study we have imposed an energy cut-off to the expected neutrino fluxes from  HESSJ1614-518 and HESSJ1616-508 sources similar to the RXJ1713.7-3946 spectrum. In case of a harder energy spectrum the discovery potential of this KM3NeT configuration to discover these sources is better. Furthermore, the discovery potential for galactic sources can be improved significantly by optimizing the geometrical parameters of the telescope's layout (e.g.~the distance between DUs) as well as developing more efficient tracking algorithms that take into account the known source direction into the signal selection and track reconstruction \cite{apostol-2}.
\section*{Acknowledgment}
The KM3NeT project is supported by the EU in FP6 under Contract 140 no.~011937 and in FP7 under Grand no.~212525.
\bibliographystyle{elsarticle-num}

\end{document}